\documentclass[twocolumn,notitlepage,floats,nofootinbib,amsmath,amssymb,aps,prd]{revtex4-1}

\usepackage{amsmath,amssymb,amsfonts,latexsym,cancel,amsthm}

\usepackage{mathtools}
\usepackage{graphicx}
\usepackage{color}
\usepackage{amsbsy}
\usepackage{hyperref}
\usepackage{tikz}
\usepackage{seqsplit}
\usepackage{comment}

\newcommand{\be}{\begin{equation}}
\newcommand{\beq}{\begin{equation}}
\newcommand{\ee}{\end{equation}}

\def\bea {\begin{eqnarray}}
\def\eea {\end{eqnarray}}

\def\dd{{\rm d}}

\definecolor{newgreen}{rgb}{0.0, 0.75, 0.0}
\definecolor{cadmiumgreen}{rgb}{0.0, 0.42, 0.24}

\begin{document}

\title{Effective LQG-inspired dynamics of a thin shell and the fate of a collapsing star}

\author{Francesco Fazzini} \email{francesco.fazzini@unb.ca}
\affiliation{Institute for Quantum Gravity, Friedrich-Alexander-Universität Erlangen-Nürnberg, \\
Staudtstr. 7, 91058 Erlangen, Germany}

\begin{abstract}
Effective models of gravitational collapse inspired by loop quantum gravity typically resolve the central singularity by replacing it with a bounce of the matter density in the Planckian regime. In the specific model analyzed here, this bounce is generally followed by the formation of shell-crossing singularities. The purpose of this work is to provide a physically meaningful extension of spacetime beyond the shell-crossing singularity. To this end, we derive the dynamics of a dust thin shell within the effective hamiltonian framework. The motion of the shell remains timelike throughout: after undergoing a quantum-gravitational bounce, it expands and eventually emerges from the white hole vacuum region.

\end{abstract}

\maketitle

\section{Introduction}

The fate of a black hole is a topic of central importance in astrophysics and quantum gravity. Einstein’s theory of gravity predicts that a black hole forms when a star collapses under its own weight, ultimately ending in a physical singularity. This result, first obtained for particular spherically symmetric initial configurations \cite{Oppenheimer:1939ue}, was later generalized by Penrose \cite{Penrose:1964wq}. Although the inclusion of pressure or the consideration of sufficiently small stellar masses can prevent black hole formation (see, e.g., \cite{chandrasekhar1931,oppenheimer1939}), once a black hole forms, its singular fate is unavoidable within the classical theory.
This picture is generally believed to be incomplete, and its completion is expected to require a quantum theory of gravity \cite{Penrose1980}. Indeed, during classical collapse, the stellar energy density eventually reaches the Planck regime, where quantum gravity effects are expected to play a dominant role.
Among the candidate theories of quantum gravity, loop quantum gravity (LQG) \cite{Rovelli:2004tv,Thiemann:2007pyv} is a leading contender. Its symmetry-reduced sectors are nowadays capable, at least at the effective level, of addressing problems of major physical interest, such as stellar collapse (see, e.g., \cite{Kelly:2020lec,Giesel:2023hys,Modesto:2006qh,Ziprick:2009nd,Tavakoli:2013rna,Rovelli:2014cta,Barrau:2014hda,Haggard:2014rza,Barcelo:2014npa,Barcelo:2014cla,Barrau:2014yka,Barrau:2015uca,Christodoulou:2016vny,Malafarina:2017csn,Olmedo:2017lvt,Bianchi:2018mml,BenAchour:2020bdt}) and cosmological evolution (for reviews see \cite{Bojowald2005,Ashtekar2009}). While the spherical symmetry reduction complicates the quantum analysis compared with the homogeneous sector (see, e.g., \cite{Husain:2022gwp}), where models are under control both at the effective and the quantum level, the effective approach remains extremely useful for studying how quantum gravity may affect gravitational collapse. It should be stressed that these effective models modify the classical setup in ways inspired by LQG, but at this stage they cannot yet be regarded as direct spherically symmetric loop quantizations.
Among the different approaches, a particularly relevant one was first introduced in \cite{Kelly:2020uwj,Kelly:2020lec}. The strength of this approach lies in the fact that it reproduces the effective dynamics of loop quantum cosmology (LQC) in the homogeneous sector while retaining the correct classical limit. Quantum gravitational effects are indeed suppressed when scalar quantities remain far from the Planck regime, which is achieved by adopting the improved dynamics scheme of LQC \cite{Ashtekar:2006wn}. Briefly, the canonical formulation is expressed in the usual Ashtekar–Barbero variables, and the modifications inspired by LQG are implemented only in the gravitational part of the hamiltonian constraint. In contrast, the gravitational part of the diffeomorphism constraint, as well as the matter contributions to both the hamiltonian and diffeomorphism constraints, are kept classical. The modification in the gravitational hamiltonian affects the curvature term, which is “holonomized” by expressing it in terms of holonomies of the extrinsic curvature, with a physical length proportional to the Planck length. This framework is often referred to as the $\bar{\mu}+K$ loop quantization scheme.

Research in this direction has in recent years focused on dust collapse, both for the Oppenheimer–Snyder model (see, e.g., \cite{Kelly:2020lec,Giesel:2023hys,Han:2023wxg,Cafaro:2024,Bobula:2023kbo,Lewandowski:2022zce,Fazzini:2023scu,Boldorini:2025dza}) and for more general profiles beyond Oppenheimer–Snyder \cite{Husain:2022gwp,Fazzini:2023ova,Bobula:2024chr}, in both the marginally and non-marginally bound cases \cite{Cipriani:2024nhx}, as well as in the presence of pressure \cite{Cafaro:2024lre}. A common feature of the dynamics studied in these works is the bouncing behavior of the stellar core when the energy density reaches the Planck regime (similarly to what occurs in LQC), due to quantum gravitational repulsion. The bounce is then followed by the formation of shell-crossing singularities (SCS).
Shell-crossing singularities occur when two matter layers composing the star intersect during evolution, leading to a divergence in the energy density. This phenomenon already appears in the classical evolution of specific initial data, but it can be avoided at the classical level through an appropriate choice of the initial density profile \cite{Hellaby:1985}. Consequently, SCS are generally considered less pathological than the central singularity: geodesic deviations remain finite near them \cite{Szekeres:1995gy}, and they do not represent a serious violation of the cosmic censorship conjecture \cite{Penrose:1999vj}. However, while SCS can be avoided classically, within LQC-inspired effective collapse they turn out to be unavoidable. This has been shown to hold generally for dust collapse \cite{Fazzini:2023ova}, and also for perfect and non-perfect fluids with pressure \cite{Cafaro:2024lre}. As a result, SCS must be regarded as central features of effective stellar collapse, much as the central singularity is in classical collapse.
Another reason why such singularities are often considered non-problematic is that, in principle, the spacetime can be extended beyond them. In the classical case, this has been achieved by considering weak solutions of the integral formulation of the evolution equations \cite{Nolan:2003wp,Lasky:2006hq}, since these equations take the form of hyperbolic conservation laws in Painlevé–Gullstrand (PG) coordinates, and shell-crossing singularities correspond to characteristic crossings of such conservation laws. After the crossing, the matter is concentrated into a non-isolated thin shell, and soon afterward the entire stellar content collapses as an isolated thin shell.
However, the integral approach has significant drawbacks: it does not allow one to impose timelike motion of the thin shell, because it implicitly assumes that the Painlevé–Gullstrand time remains continuous across the shell. This assumption has been shown to be inconsistent with requiring subluminal motion for ordinary matter, and therefore, in this precise sense, unphysical \cite{Fazzini:2025zrq}. Similar problems arise when the integral approach is applied to the effective case: the thin shell generated by the crossing of physical layers generally moves in a spacelike manner, with the precise dynamics depending on the specific formulation adopted \cite{Husain:2022gwp,Fazzini:2025hsf,Liu:2025fil}. It is also worth noting that alternative approaches not based on weak solutions suffer from the same physical inconsistencies \cite{Sahlmann:2025fde}.
A possible way to overcome this superluminal issue is to study the effective thin shell dynamics within the Israel framework (see, e.g., \cite{Israel1966,Poisson2004,Berezin1990,Ansoldi2000,Crisostomo:2003bb,HajicekKijowski1997}), which guarantees timelike behavior. The aim of the present work is precisely to extend the Israel approach to the effective case and to investigate how a thin shell of matter evolves in the effective spacetime. The main result is the derivation of an effective dynamics for the dust thin shell that is always timelike, that admits the correct classical Israel limit, and that provides an approximation of the spacetime evolution after the formation of SCS.

\section{The first Israel junction condition\label{section2}}

To describe the spacetime generated by the thin shell, we assume that our four-dimensional spacetime is divided by the shell into two distinct regions: a Minkowski interior and an exterior described by the effective Schwarzschild line element in Painlevé-Gullstrand coordinates \cite{Kelly:2020uwj}:

\begin{align}
    & \dd s_-^2=-\dd t_-^2 +\dd r_-^2 +r_-^2\dd \Omega_-^2  \label{primo} ~,  \\
    &\dd s_+^2=-F_+(r) \dd t_+^2 +2 N^r_+ \dd r_+\dd t_+ +\dd r_+^2 +r_+^2 \dd \Omega_+^2 ~, \label{secondo}
\end{align}
where 

\begin{equation}
F_+=1-\frac{R_S}{r_+}+\frac{\gamma^2 \Delta R_S^2}{r_+^4} ~, \quad N^r_+=\sqrt{\frac{R_S}{r_+}\left(1-\frac{\gamma^2 \Delta R_S}{r_+^3} \right)}  ~. \label{sfarfunctions}
\end{equation}

Here $\Delta$ is proportional to the Planck area and represents the minimum non-zero eigenvalue of the area operator in loop quantum gravity, while $\gamma$ is the Barbero–Immirzi parameter, and $R_S=2GM$ is the Schwarzschild radius, with $M$ the Schwarzschild mass. The exterior is a solution of the effective hamiltonian constraint in the areal gauge, and it correctly reproduces the classical Schwarzschild solution in the regime $\Delta/r^2 \ll 1$. An important feature of this line element is the existence of two Killing horizons, determined by the zeros of $F_+$. For macroscopic black holes, the outer horizon is located approximately at $R_S$, whereas the inner horizon lies deep in the quantum regime. The maximal extension of this effective spacetime yields a Penrose diagram closely resembling that of Reissner–Nordström spacetime \cite{Lewandowski:2022zce}.

Allowing the external and internal times and radial coordinates to differ enables the implementation of the first Israel junction condition. Considering the chart ${\tau, \theta, \phi}$ on the shell hypersurface, with $\tau$ the shell proper time, the induced metric on the shell reads

\begin{equation}
 \dd s^2=-\dd \tau^2 +R(\tau)^2 \dd \Omega^2 ~. \label{1} 
\end{equation}

Following Israel’s approach, we require that this induced metric be identical when computed from the interior and exterior spacetimes, thereby ensuring the continuity of the induced metric:

\begin{align}
  &  \dd s^2_-=-\left(\Dot{T}_-^2-\Dot{R}_-^2 \right)\dd \tau^2+ R_-(\tau)^2 \dd \Omega_-^2  \label{2} ~, \\
  &\dd s^2_+=-\left(F_+\Dot{T}_+^2-2N^r_+ \Dot{T}_+ \Dot{R}_+  -\Dot{R}_+^2 \right)\dd \tau^2+ R_+(\tau)^2 \dd \Omega_+^2  \label{3}~,
  \end{align}
where $r_\pm(\tau)\equiv R_\pm(\tau)$, $t_\pm(\tau)\equiv T_\pm(\tau)$, and the dot denotes differentiation with respect to the shell proper time.

The first Israel junction condition then requires the equivalence of \eqref{1}, \eqref{2}, and \eqref{3}, leading to

\begin{equation}
    \begin{cases}
        & R_-(\tau)=R_+(\tau)=R(\tau) ,~~ \theta_-=\theta_+=\theta, ~~ \phi_-=\phi_+=\phi   \\
        &   \Dot{T}_-^2-\Dot{R}_-^2=1 \\
        & F_+ \Dot{T}_+^2-2N^r_+\Dot{T}_+ \Dot{R}_+-\Dot{R}_+^2=1    
    \end{cases}
    \label{a}
\end{equation}

The second and third conditions ensure that $\tau$ represents the proper time of the shell as measured by \emph{both} metrics, and guarantee that the shell motion is timelike throughout its evolution. The can be explicitly demonstrated by constructing the shell four-velocity using both the interior and exterior metrics

\begin{align}
      u^{\mu}_\pm=\left\{\frac{\dd T_\pm}{\dd \tau}, ~\frac{\dd R}{\dd \tau}, ~0, ~0 \right\}  ~.
    \end{align}

and requiring $u^\mu_{\pm}u_{\mu \pm}=-1$, leads to the last two conditions in \eqref{a}. Subluminal motion is not ensured by simply equating the $\tau$-$\tau$ components of \eqref{2} and \eqref{3}.

To proceed with the Israel approach, we need to relate the discontinuity in the extrinsic curvature across the shell to the surface stress–energy tensor of the thin shell, which leads to (violation of) the second Israel junction condition. The Israel formalism for thin shells in general relativity is usually developed within the lagrangian formulation of Einstein’s theory. However, effective models of gravitational collapse inspired by loop quantum gravity are more naturally expressed in the canonical framework, where the connection with the underlying fundamental theory is more transparent. The hamiltonian formulation of the junction conditions has been extensively studied at the classical level \cite{Ansoldi2000,Crisostomo:2003bb,HajicekKijowski1997}, and by adapting these techniques we will derive the effective dynamical equation for the thin shell.

\section{Hamiltonian formulation of the second Israel junction condition}

The basic idea behind the canonical approach to the second Israel junction condition is to solve the integrated version of the constraints across the thin shell, located in $R$

\begin{align}
& \int_{R-\varepsilon}^{R+\varepsilon}  N (\mathcal{H}^{grav.}_t+ \mathcal{H}^{matt.}_t )\dd r\approx 0 \\
&\int_{R-\varepsilon}^{R+\varepsilon} N^r (\mathcal{H}^{grav.}_r+\mathcal{H}^{matt.}_r) \dd r \approx 0 ~, \label{diffeoconstraint}
\end{align}

taking the limit $\varepsilon \rightarrow 0$ after integration. This allows one to solve the constraint in a distributional sense, since, as we will see shortly, $\mathcal{H}^{matt.}_t $ and $\mathcal{H}^{matt.}_r$ are distribution-valued. Here, 
$R$ at the integration boundaries corresponds to the shell location at a given coordinate time $t$, which can in general differ between the interior and the exterior regions. We can further simplify the expression by imposing the areal gauge. To do so, we first need to express the gravitational constraints in terms of the Ashtekar–Barbero variables, which is convenient for transitioning to the effective theory. In the classical theory, the two constraints read

\begin{align}
\mathcal{H}^{grav.}_t=&-\frac{1}{2G\gamma}\bigg[ \frac{2ab \sqrt{E^a}}{\gamma}+\frac{E^b}{\gamma \sqrt{E^a}}(b^2+\gamma^2)-\notag \\
&-\frac{\gamma (\partial_r (E^a)^2)}{4 E^b \sqrt{E^a}}-\gamma \sqrt{E^a}\partial_r \left(\frac{\partial_r E^a}{E^b}\right) \bigg]~,
\end{align}
and 
\begin{equation}
\mathcal{H}^{grav.}_r=\frac{1}{2G \gamma}(2E^b \partial_r b-a \partial_r E^a) ~,
\end{equation}

where $a$ and $b$ denote the radial and angular components of the extrinsic curvature (up to a $\gamma$ factor), and $E^a$ and $E^b$ correspond to the radial and tangential components of the densitized triads \cite{Kelly:2020uwj}. Note that the superscripts $a$ and $b$ do not run, but rather label the only non-trivial components of the densitized triad. The areal gauge is fixed by setting $E^a=r^2$, since the general line element in this formulation reads

\begin{equation}
\dd s^2=-N^2 \dd t^2 +\frac{(E^b)^2}{E^a}\left(\dd r + N^r \dd t \right)^2+E^a \dd \Omega^2 ~,
\end{equation}

and imposing the vanishing of the radial diffeomorphism constraint $\mathcal{H}^{matt.}_r+ \mathcal{H}^{grav.}_r \approx 0$, gives

\begin{equation}
 a=E^b \partial_r b +\frac{G \gamma}{r}\mathcal{H}^{matt.}_r ~.   
\end{equation}
Notice that this solves automatically \eqref{diffeoconstraint}.
Substituting the previous expression in the gravitational part of the hamiltonian constraint, and using the fact that both the solutions \eqref{1} and \eqref{2} have $E^b=r$, gives, after a bit of computations

\begin{align}
\mathcal{H}^{grav.}_t=&-\frac{1}{2G\gamma^2}\partial_r(rb^2)-\frac{b}{\gamma}\mathcal{H}^{matt.}_r ~= \notag \\
=&-\frac{1}{2G\gamma^2}\partial_r(rb^2)+N^r \mathcal{H}^{matt.}_r ~,
\end{align}

diwhere the relation between the shift vector and the $b$ field can be easily derived by imposing that the areal gauge is preserved during the dynamics \cite{Kelly:2020uwj}.
The polymerization of this constraint proceeds in the same way as in the cases of stellar collapse and vacuum \cite{Kelly:2020uwj,Husain:2022gwp}: the $b$ variable in the first term is polymerized through

\begin{equation}
    b \rightarrow \frac{r}{\sqrt{\Delta}}\sin \left(\frac{\sqrt{\Delta}b}{r} \right) ~,
\end{equation}

while for the shift vector 

\begin{equation}
 N^r \rightarrow -\frac{r}{2\gamma \sqrt{\Delta}}\sin\left(\frac{2\sqrt{\Delta}b}{r} \right) ~. 
\end{equation}

This polymerization of the shift vector, after imposing the areal gauge, has been shown to be consistent with the underlying effective theory which preserves (spatial) diffeomorphism invariance \cite{Giesel:2023hys}.
To be consistent with the line elements \eqref{primo} and \eqref{secondo}, we can fix the lapse to $N=1$. Note however that the shift $N^r$ exhibits a discontinuity across the shell surface, located at $R$. With this choice, the equation we need to solve reduces to

\begin{equation}
\int_{R-\varepsilon}^{R+\varepsilon} \left[  \mathcal{H}^{grav.}_t+ \mathcal{H}^{matt.}_t   \right] \dd r \approx 0 \label{fundamental} ~,
\end{equation}

where the weak equality holds on shell. To proceed further, we need to write the constraints explicitly in this gauge. Let us start with the scalar (hamiltonian) constraint for gravity, which reads

\begin{align}
  \mathcal{H}^{grav.}_t=&-\frac{1}{2G\gamma}\left[\frac{1}{\gamma}\partial_r \left(\frac{r^3}{\Delta} \sin^2\frac{\sqrt{\Delta}b}{r} \right)\right]+ \notag \\ 
  &+N^r \mathcal{H}^{matt.}_r
  =\Tilde{\mathcal{H}}^{grav.}_t+N^r \mathcal{H}^{matt.}_r
  \label{hamilton} ~.
\end{align}

The integral of $\Tilde{\mathcal{H}}^{grav.}_t$ can be computed straightforwardly. To this end, we recall the solutions for the interior and exterior in terms of the $b$ field \cite{Husain:2022gwp}

\begin{align}
 &b^-(r,t)=0 ~,\\
 &b^+(r,t)=-\frac{r}{\sqrt{\Delta}}\sin^{-1}\sqrt{\frac{2G M\gamma^2 \Delta}{r^3}} ~,
\end{align}

where $M$ is the mass parameter appearing in the effective Schwarzschild solution. By integrating the first term in \eqref{hamilton}, we obtain boundary terms (recall that the areal radius is continuous across the shell, as ensured by the first Israel junction condition)

\begin{equation}
\int_{R-\varepsilon}^{R+\varepsilon} \Tilde{\mathcal{H}}_t^{grav.}\dd r= - \frac{r^3}{\Delta}\sin^2 \frac{\sqrt{\Delta}~b}{r}\bigg|_{R-\varepsilon}^{R+\varepsilon} = -M  ~, \label{veryimpo}
\end{equation}
where the last equality holds once the limit $\varepsilon\rightarrow 0$ is performed.
Let us now focus on the matter part of the constraints. In general, the energy–momentum tensor for a dust thin shell is given by

\begin{equation}
    T^{\pm}_{\mu \nu}=\sigma u^\pm_\mu u^\pm_\nu  \delta(\chi_\pm) ~,
\end{equation}

where $u^\mu_\pm$ is the shell $4$-velocity as measured by the interior and exterior, $\sigma$ is the proper surface energy density, and $\dd \chi_\pm$ represents the infinitesimal proper distance in the shell comoving frame. We will later discuss the relation between the coordinates $\chi$ and $r$. 

Since the shell energy--momentum tensor must be expressed in the frame adapted to the spacetime foliation defined by the $\{t^{\pm}, r, \theta, \phi\}$ coordinates, it is necessary to project it along the PG time direction. To this end, we introduce the $4$-vector $n^{\mu}$ generating the PG-time flow

\begin{align}
   & n^{\pm}_\mu=\left\{-1,~0,~0,~0 \right\}~, \\
   & n^{\mu}_+=\left\{1,-N^r_+,~0,~0 \right\}~, \quad n^\mu_-=\left\{1,~0,~0,~0 \right\}~.
   \end{align}

Then, projecting $T_{\mu \nu}$ along $n^\pm_\mu$ gives the matter part of the scalar constraint

\begin{align}
\mathcal{H}^{matt.}_t=&4 \pi R^2 T_{\perp\perp}= 4 \pi R^2 n^{\mu} n^\nu u_\mu u_\nu \sigma \delta(\chi) = \notag \\
=&4 \pi R^2(n \cdot u)^2 \sigma \delta(\chi) ~,  \label{hmatter0} 
\end{align}
which formally holds both for the interior and exterior. A direct computation gives $(n_\pm \cdot u_\pm )^2=\Dot{T}^2_\pm $. One can then impose the first Israel junction conditions to the shell $4$-velocity to relate these expressions with the radial component of the $4$-velocity

\begin{align}
 &\Dot{T}_+=\frac{\Dot{R} N^r_+ +\sqrt{\Dot{R}^2+F_+}}{F_+} \label{t+} ~, \\
 & \Dot{T}_-=\sqrt{\Dot{R}^2+1}  \label{t-} ~,
\end{align}

where $F_+$ and $N^r_+$ are given, respectively, by the first and second of \eqref{sfarfunctions}, evaluated at $r=R(\tau)$. Notice that, as for the gravitational part of the scalar constraint, $\mathcal{H}_{t}^{matt.}$ exhibits a discontinuity across the thin shell. 

To complete the computation and integrate properly along the $r$ direction, we need to express the delta function in terms of the areal radius $r$. To this end, we must find the normalized spacelike $4$-vector orthogonal to $u^\mu$ and perpendicular to the $2$-dimensional spacelike hypersurface describing the shell location, which provides the proper distance $\dd \chi$. These conditions can be written mathematically as follows:

\begin{equation}
    \begin{cases}
      & \chi^\mu_\pm \chi^{\pm}_\mu=1    \label{chisei} \\
      & \chi^\mu_\pm u^\pm_\mu=0 \\
      & \chi^\mu_\pm= \{A_\pm(\tau),B_\pm(\tau), 0, 0\}
    \end{cases}
\end{equation}

The last condition ensures that the $4$-vector has no components along the tangential directions to the shell. A direct computation leads to 
\begin{align}
&\chi^\mu_-=\left\{\Dot{R}, \Dot{T}_-,0,0 \right\} ~, \label{chi-} \\
& \chi^\mu_+= \left\{ \frac{\Dot{R}+N^r_+ \sqrt{\Dot{R}^2+F_+}}{F_+}, \sqrt{\Dot{R}^2+F_+},0,0   \right\}\label{chi+} ~.
\end{align}

It is easy to verify that $\dd \chi $ gives the proper radial distance in the shell comoving frame. Given the displacement vector $\chi^\mu \dd \chi$ along the direction individuated by $\chi^\mu$, we have

\begin{align}
 g_{\mu \nu} \chi^\mu \chi^\nu \dd \chi^2=&(h_{\mu \nu}-u_\mu u_\nu)\chi^\mu \chi^\nu \dd \chi^2 = \\
 =& h_{\mu \nu} \chi^\mu \chi^\nu \dd \chi^2= \dd s^2=\dd \chi^2 ~,
\end{align}

where $\dd s^2$ is the squared radial proper distance in the shell comoving frame, since $h_{\mu\nu}$ is the projection of the metric in the submanifold orthogonal to shell $4$-velocity. In the last step, we used the first of \eqref{chisei}. The previous computation formally holds for both the interior and exterior regions. 

Now, to express the Dirac delta in terms of the coordinate $r$, we need the transformation law between the two charts $\{t, r, \theta, \phi\}$ and $\{\tau, \chi, \theta, \phi\}$. It is given by

\begin{equation}
    \begin{cases}
        & \dd t = \frac{\partial t }{\partial \tau}\big|_{\chi} \dd \tau + \frac{\partial t}{\partial \chi}\big|_\tau \dd \chi =u^t \dd \tau + \chi^t \dd \chi    \\
        & \dd r = \frac{\partial r }{\partial \tau}\big|_{\chi} \dd \tau + \frac{\partial t}{\partial \chi}\big|_\tau \dd \chi =u^r \dd \tau + \chi^r \dd \chi
        \end{cases} \label{system}
\end{equation}
where the $\pm$ label is omitted for brevity, but this transformation is in general different between the interior and exterior. 

The integration of the scalar constraint (as well as the diffeomorphism constraint) is carried out along the $t=\text{const.}$ hypersurface, which is in general discontinuous across the shell (see \cite{Fazzini:2025zrq}). Therefore, from the previous equations we have

\begin{equation}
 \frac{\dd r}{\dd \chi_\pm}=-u^r_\pm \frac{\chi^t_\pm}{u^t_\pm}+\chi^r_\pm  =\frac{1}{\Dot{T}_\pm} ~.
\end{equation}

The transformation of the Dirac delta follows:
\[
\delta(\chi_\pm) = \frac{1}{\Dot{T}_\pm} \, \delta(r - R(t_\pm))~,
\]
where, as before, $R(t_\pm)$ is the shell location at times $t_\pm$, which in general differ from each other. By substituting these results into \eqref{hmatter0}, and using \eqref{t+} and \eqref{t-} respectively for the exterior and interior, we obtain

\begin{align}
 & \mathcal{H}^{matt.}_{t+}= 4 \pi R^2 \sigma~ \frac{\Dot{R}N^r_+ +\sqrt{\Dot{R}^2+F_+}}{F_+} ~\delta(r-R) ,~for~~ r\geq R  \\
 &\mathcal{H}^{matt.}_{t-}= 4 \pi R^2\sigma\sqrt{1+\Dot{R}^2}~\delta(r-R) ~, ~for ~~r<R ~.
 \end{align}

What remains to complete the computation is the matter part of the diffeomorphism constraint. To compute it, we need one of its indices projected along the normal to the spatial slice, and the other along the spatial radial direction \cite{Crisostomo:2003bb}. This can be written as

\begin{align}
  \mathcal{H}^{matt.}_{r}=& 4 \pi R^2 T_{\perp r}=4 \pi R^2 u_\nu u_\nu n^\mu h^{\nu}_r \sigma \delta(\chi)= \notag\\
  =&4 \pi R^2 (n \cdot u) u_r \sigma \delta (\chi) ~,
\end{align}

where $h^\nu_r = \delta^\nu_r + n^\nu n_r$. We can then transform the Dirac delta as was done for the scalar constraint, expressing it in terms of $r$. After a brief computation, we obtain

\begin{equation}
\mathcal{H}^{matt.}_{r}=-2 \pi R^2 \sigma \frac{\Dot{R}+N_+^r \sqrt{\Dot{R}^2+F_+}}{F_+}\delta(r-R) ~,
\end{equation}

for $r\geq R$, while for $r < R$ the matter part of the diffeomorphism constraint gives no contribution to \eqref{fundamental}, since $N^r_- = 0$. By collecting all the results, we are now in a position to compute \eqref{fundamental}, which explicitly reads

\begin{equation}
 M= 2 \pi \sigma R^2 \left( \sqrt{\Dot{R}^2+1}+\sqrt{\Dot{R}^2+1-\frac{R_S}{R}+\frac{\gamma^2 \Delta R_S^2}{R^4}}\right) \label{semifinal} ~. 
\end{equation}

To further simplify this expression, we recall that in the effective theory the matter part of the Einstein equations remains unchanged, and therefore the covariant conservation of the energy-momentum tensor still holds. As in the classical case, if the interior and exterior regions are vacuum, $\nabla_\mu T^{\mu \nu} = 0$ can be integrated across the shell, leading to $D_a S^a_b = 0$, where $a,b = \tau, \theta, \phi$ and $S^a_b = diag\{-\sigma, 0, 0\}$, and the covariant derivative is taken with respect to the induced metric on the shell \eqref{1}. A straightforward explicit computation then leads to

\begin{equation}
    \partial_\tau (4\pi R^2 \sigma)=0 ~,
\end{equation}

which means  

\begin{equation}
    4 \pi R^2 \sigma=const.\equiv m ~,
\end{equation}

where $m$ is the inertial mass of the shell, since $\sigma$ is the proper surface density. We can insert the previous in \eqref{semifinal}, to get

\begin{equation}
    M=\frac{m}{2}\left(\sqrt{\Dot{R}^2+1}+\sqrt{\Dot{R}^2+1-\frac{R_S}{R}+\frac{\gamma^2 \Delta R_S^2}{R^4}}\right) ~. \label{fondamentale}
\end{equation}

The classical limit is recovered for $R^2 \gg \Delta$, which can be translated in terms of the energy density as $\sigma \ll m / (4 \pi \Delta)$. Notice also that the classical limit of \eqref{fondamentale} coincides with the equation usually derived using the Schwarzschild vacuum in Schwarzschild coordinates for the exterior. The two expressions are equivalent, since both the PG coordinates and the Schwarzschild coordinates employ the areal gauge, and the time $T^+$ does not appear in the previous expression; $T$ generally differs between the two coordinate systems. 

The reason why this result is derived in PG coordinates for the exterior is that the diffeomorphism algebra is deformed at the effective level, and writing \eqref{3} in Schwarzschild coordinates does not satisfy the hamiltonian constraint. For a discussion on this, see \cite{Giesel:2023hys}.

\section{The effective dynamics for the thin shell}

The result obtained in the previous section constitutes the main result of this work. Before analyzing its properties and solutions, a few comments are in order. First, this approach provides a hamiltonian formulation of the Israel junction conditions, and for this reason the shell dynamics is timelike at all times, with $\tau$ being the shell proper time. This result is remarkable, since alternative approaches to the same problem developed so far do not enforce subluminal motion of the shell—whether measured with respect to at least one side of the spacetime when the metric is discontinuous across the shell \cite{Husain:2022gwp,Fazzini:2025hsf}, or with respect to the continuous metric \cite{Liu:2025fil,Sahlmann:2025fde}.  

Another important feature to notice is that the value of $m$ determines the shell kinetic energy at infinity. For $m=M$ (the marginally bound case), at $R=+\infty$ we have $\Dot{R}=0$ (the collapse starts at infinity with zero kinetic energy). Since in the classical limit we recover the Israel dynamics, we can also infer that for $m<M$ the shell will start at $R<+\infty$ with zero kinetic energy (bound case), while for $m>M$ it starts at $R=+\infty$ with non-zero kinetic energy (unbound case). In this work, we are interested in the marginally bound case, and from now on we proceed with the condition $m=M$. Comments on the other cases are given at the end of this section.  

We can then manipulate the equation of motion to extract quantitative features. By isolating the first term on the right-hand side and squaring \eqref{fondamentale}, after a few steps we obtain

\begin{equation}
\left(\frac{\Dot{R}}{R}\right)^2=\frac{R_S}{2R^3}\left(1-\frac{\gamma^2 \Delta R_S}{R^3} \right)\left[1+\frac{R_S}{8R}\left(1-\frac{\gamma^2 \Delta R_S}{R^3} \right) \right] ~. \label{beautifull}
\end{equation}

In this formulation is manifest the role of the effective corrections. We have two zeroes for the right-hand side, in

\begin{align}
 &R=(\gamma^2 \Delta R_S)^{\frac{1}{3}} ~,  \label{primaeq}    \\
 & R^3 R_S +8 R^4= \gamma^2 \Delta R_S^2  ~. \label{secondaeq}
\end{align}

It is, however, easy to check that the second condition is never attained during the dynamics for a shell starting at 
$R > (\gamma^2 \Delta R_S)^{\frac{1}{3}}$, since the solution of \eqref{secondaeq} is larger than that of \eqref{primaeq}, and \eqref{primaeq} represents a turning point for the solution. This statement can be easily verified by studying the equation for $R \sim (\gamma^2 \Delta R_S)^{\frac{1}{3}}$. In this range, the second term in \eqref{beautifull} becomes negligible, and therefore

\begin{equation}
    \left(\frac{\Dot{R}}{R}\right)^2\sim \frac{R_S}{2R^3}\left(1-\frac{\gamma^2 \Delta R_S}{R^3} \right)~. 
    \end{equation}

The previous equation has an analytic solution, given by

\begin{equation}
R(\tau)=\left( \frac{9}{8} R_S \tau^2 +\gamma^2\Delta R_S\right)^{\frac{1}{3}} ~, \label{solution}
\end{equation}

where we assumed the bouncing time as the initial time. It is manifest that this solution describes a bouncing shell. The surface proper energy density at the bounce is
\begin{equation}
\sigma(R_{ bounce}) = \frac{M^{\frac{1}{3}}}{4 \pi (2 G \gamma^2 \Delta)^{\frac{2}{3}}}.
\end{equation}

We can compute the average energy density $\Bar{\rho}$ contained in the spherical volume of radius $R_{bounce}$, through

\begin{equation}
  \bar{\rho}_{bounce}=\frac{\sigma A_{bounce}}{V_{bounce}}=\frac{3}{8 \pi G \gamma^2 \Delta}=\rho_{crit.}  ~,
\end{equation}

showing that the planckian upper bound of the volumetric energy density for continuous profiles here is respected in average.

A few additional comments are in order. First, the bouncing radius of the thin shell in the marginally bound case is exactly the same as in the Oppenheimer-Snyder effective model \cite{Kelly:2020lec,Lewandowski:2022zce,Bobula:2023kbo,Fazzini:2023scu,Giesel:2023hys}. This implies, in particular, that the shell bounces in the non-trapped region of the effective vacuum spacetime and clearly moves toward the antitrapped region, as in the Oppenheimer-Snyder picture \cite{Lewandowski:2022zce,Bobula:2023kbo,Fazzini:2023scu,Giesel:2023hys}. This behavior is expected from the fact that the thin shell motion is timelike \cite{Fazzini:2025zrq}, but here it is explicitly verified. This result contrasts with other works in the literature \cite{Husain:2022gwp,Fazzini:2025hsf,Liu:2025fil,Sahlmann:2025fde}, where the thin shell is allowed to be outgoing within the black hole trapped region, leading to superluminal motion. On the other hand, the dynamics considered here is similar to that of a classical charged thin shell \cite{DeLaCruz1967}, in which the bounce is caused by the repulsive character of the Reissner–Nordström spacetime close to $r=0$.

It is important to notice that the solution \eqref{solution} is expressed in the shell proper time, not in the exterior or interior PG times (even though such a transformation is possible). Therefore, even if the solution closely resembles the Oppenheimer-Snyder one in terms of the stellar radius, it actually differs: in the Oppenheimer-Snyder case, the time is the dust proper time, which coincides with the coordinate time of the interior written in PG (or equivalently LTB) coordinates \cite{Giesel:2023hys}, and with the exterior PG time until the bounce \cite{Fazzini:2023scu}. Here, $\tau$ is also the dust proper time, but it does not coincide with either the exterior or interior PG times; it is related to them via the first of \eqref{system}.  

By examining \eqref{t+} closely, we see that the advanced Painlevé-Gullstrand time for the exterior does not cover the entire shell dynamics. While the equation behaves well in the pre-bounce phase, as the shell approaches the inner white hole horizon ($1 - (N^r_+)^2 \rightarrow 0$, $\Dot{R} > 0$), the right-hand side diverges. This indicates that an external PG observer sees the shell approaching that Killing horizon in infinite PG time, consistent with the fact that the external PG chart does not cover the maximal vacuum extension, as in the Reissner-Nördstrom or Kruskal spacetimes, and is therefore not a suitable coordinate chart for describing the post-bounce evolution. A similar behavior has been found in the effective Oppenheimer-Snyder case \cite{Fazzini:2023scu}. In contrast, the shell proper time is well-defined throughout the entire evolution.  

A further comment is needed for cases beyond the marginally bound one. In the gravitational collapse of a continuous dust profile, it has been found—both in the Oppenheimer-Snyder case \cite{Cafaro:2024} and for more general profiles \cite{Cipriani:2024nhx}—that the vacuum exterior depends on the spatial curvature of the outermost shell of the distribution in order to ensure metric continuity, since the Birkhoff theorem does not hold at the effective level. In particular, the effective vacuum presented here acquires terms proportional to the spatial curvature, and different vacua are not diffeomorphic to each other, even though they all converge to the classical Schwarzschild solution in the classical limit. For this reason, it is not obvious that, for $m > M$ or $m < M$, the correct exterior is the one given by \eqref{secondo}, with $F_+$ and $N^r_+$ as in \eqref{sfarfunctions}.

To conclude, we make a comment regarding the covariance of the model. The effective theory considered in this work has a deformed covariance, in the sense that the constraint algebra closes, but the timelike diffeomorphisms generated by the constraints are not exactly classical. In particular, it is possible (although this needs to be explicitly checked) that the coordinate transformation leading to the shell proper time causes the metric to transform in a non-classical way, making the imposition of condition \eqref{3} formally incorrect. If this turns out to be the case (which will be the subject of future work), then the validity of this shell collapse model would still be preserved in the covariant theory with the same vacuum solutions as the present one, provided that the integral over the radial coordinate of the hamiltonian constraint still yields the Schwarzschild mass on shell, as in \eqref{veryimpo}. This is the case, for example, in the covariant model of \cite{Alonso-Bardaji:2025hda}, where one can carry out the same computation presented here, ending up with the same result.

\section{Implication for effective stellar collapse}

Even though the results obtained in the previous section are limited to the collapse of a dust thin shell, they can be used to approximate part of the dynamics in the stellar case. As mentioned in the introduction, effective stellar collapse predicts a bounce of the core when the energy density reaches the Planck scale, followed by the formation of shell-crossing singularities (SCS). A shell-crossing singularity can be interpreted as a non-isolated thin shell, since at an SCS the gravitational mass exhibits a jump discontinuity \cite{Husain:2022gwp,Cipriani:2024nhx,Fazzini:2025hsf}. As the thin shell forms, the gravitational mass in the stellar core is pushed outward by repulsive gravity, while the stellar tail is pushed inward as it still evolves in the black hole trapped region (having not bounced yet). Consequently, the non-isolated thin shell rapidly acquires the entire mass content of the original star, effectively becoming an isolated thin shell. This picture is supported by numerical simulations of weak solutions, where the solutions inside and outside the shell are equivalent to the solution of the original PDE (for a discussion, see \cite{Fazzini:2025hsf}). Therefore, we can approximate the post-bounce dynamics with that of a thin shell carrying the gravitational mass of the original star, as given by the solution of \eqref{beautifull}. The motion remains timelike throughout, and the expanding thin shell eventually emerges from the white hole outer horizon of the maximal extension of \eqref{2}.

In this picture, the resulting exterior is very similar to that provided by the Oppenheimer-Snyder collapse, although the interior dynamics is drastically different. In the Oppenheimer-Snyder case, shell-crossing singularities do not occur \cite{Fazzini:2023ova}, and the matter dynamics is symmetric under time reversal around the bounce point. Here, the post-bounce dynamics differs significantly from the pre-bounce evolution and is dominated by the evolving thin shell.  

It is also noteworthy that the qualitative post-bounce dynamics is similar for very different initial energy density profiles, provided they develop shell-crossing singularities either before or after the bounce. This holds for any initial energy density profile with compact support, as well as for non-compact profiles with sufficiently large inhomogeneities, as shown in \cite{Fazzini:2023ova}. This implies that extracting information about the original collapsed star by observing the expanding thin shell could be challenging. This issue is expected to be mitigated by the inclusion of pressure (either matter pressure or quantum gravitational effects), which can reduce or prevent shell-crossing singularities and thus avoid thin shell formation. From this perspective, the model considered here should be regarded as a toy model for effective stellar collapse.

\section{Conclusions}

The model presented in this work concerns the gravitational collapse of a marginally bound dust thin shell within an effective framework inspired by loop quantum gravity. By implementing the Israel junction conditions from a hamiltonian perspective, we have derived the effective equations governing the dust thin shell. The most relevant feature of this effective dynamics is that the collapsing shell halts its motion when its surface density becomes Planckian, bounces, and subsequently follows a dynamics symmetric under time reversal around the bounce point, similar to what occurs in the effective Oppenheimer-Snyder scenario. By construction, the shell dynamics is always timelike, the induced metric on the shell surface is continuous, and the shell proper time is well defined at all times—unlike in other approaches, such as the integral method. The effective thin shell dynamics can therefore be used to model the post-bounce stellar collapse, in particular the dust evolution beyond shell-crossing singularity formation. Since shell-crossing singularities are ubiquitous in effective dust collapse, and can be interpreted as non-isolated thin shells that rapidly acquire the entire stellar mass, studying the post-bounce dynamics through the thin shell formalism is physically justified. This implies that the entire matter content of the star emerges in another universe, passing through the anti-trapped region of the maximal extension of the effective Schwarzschild metric, in the form of an expanding thin shell.

Although the model constructed in this work seems physically sound, it leaves many questions unanswered. First, it is not clear how the known white hole instability could be avoided. Moreover, since the maximal vacuum extension of this spacetime is structurally very similar to the classical Reissner-Nördstrom solution, it may potentially be subject to the same issues, such as the mass inflation problem. These problems could, in principle, be mitigated by gluing the two asymptotic regions of the vacuum exterior, similar to the Oppenheimer-Snyder case \cite{Han:2023wxg}. If such gluing is not performed, addressing the information paradox becomes challenging, as an observer in the same universe as the collapsing star would see an almost classical evaporating black hole. These crucial issues are left for future investigation.

Finally, it is possible that additional quantum gravitational effects should be included in the constraints, such as dispersion or diffusion effects. These could prevent the formation of shell-crossing singularities and modify the external vacuum structure, potentially allowing the matter content to emerge from the first asymptotic region. Exploration of these more esotic scenarios is left for future work.

\acknowledgments

I would like to thank Hongguang Liu for helpful discussions in the preliminary stage of this work, and Edward Wilson-Ewing for helpful comments.

\end{document}